\documentstyle[sprocl,epsfig]{article}

\def\sst{\scriptscriptstyle}

\def\be{\begin{equation}}
\def\ee{\end{equation}}
\def\bea{\begin{eqnarray}}
\def\eea{\end{eqnarray}}
\newcommand{\bm}[1]{\mbox{\boldmath $#1$}}

\begin{document}

\title{
\begin{flushright}
VUTH 98-17\\
LANCS/HEP/9806
\end{flushright}
\mbox{} \\
SPIN PHYSICS IN DEEP INELASTIC SCATTERING \footnote{
Summary talk of Spin Physics working group at the
$6^{th}$ International Workshop on Deep Inelastic Scattering
and QCD, Brussels (1998).}
}

\author{P.J. MULDERS}
\address{Physics Department, Free University,\\
De Boelelaan 1081, 1081 HV Amsterdam, the Netherlands\\
E-mail: mulders@nat.vu.nl}

\author{T. SLOAN}
\address{University of Lancaster\\
Lancaster   LA1 4YB, UK\\
E-mail: t.sloan@lancaster.ac.uk}

\maketitle

\abstracts{
The problem of our understanding of the spin  structure of the nucleon has
been with us since the publication of the EMC measurements of the polarised
structure function of the proton in 1987.  In this talk a brief history of
the subject is given followed by a review of the results
presented in this workshop and the progress made to date.
}

\section{Introduction}

In the simple quark model the  spin of the proton is carried by
its three valence quarks so that $\Delta\Sigma= \Delta u + \Delta d = 1$.
Here $\Delta q$ = $\int^1_0 dx\, \left(q_R(x)-q_L(x)\right)$   where
$q_{R(L)}(x)$ are quark distributions for chirally right\-handed 
(left\-handed) quarks 
in a polarized proton and $q$ = $u, d$, etc. indicates the quark flavours.  
In a fast-moving proton the $q_{R(L)}$ are associated with quarks spinning 
parallel (antiparallel) to the proton helicity.
The simple quark model is clearly inadequate since
it predicts that the ratio of the axial vector to vector coupling
constants in neutron $\beta$ decay is $g_A = 5/3$ compared to the
measured value of 1.26.  Relativistic effects are invoked to account for
this difference and correspondingly
reduce  $\Delta\Sigma$ to $\sim 0.7$  with the excess spin appearing as
the effects of orbital motion.

Deep inelastic scattering (DIS) with polarised
charged leptons on polarised targets allows the quark distributions
$q_{R(L)}$  to be investigated.  These are extracted from 
the structure function $g_1(x,Q^2)$ measured in polarised DIS 
using the parton model relation
$g_1(x,Q^2) =\frac{1}{2}\sum_q  e_q^2\,\left(q_R(x)-q_L(x)\right)$.
The difference in cross sections for different orientations
of the lepton and target nucleon polarisations are related to
the polarised structure functions $g_1$ and $g_2$ by:
\begin{equation}
d\sigma^{\uparrow\downarrow}-d\sigma^{\uparrow\uparrow}=
a g_1(x,Q^2) + b g_2(x,Q^2)
\label{asym}
\end{equation}
where $a$ and $b$ are kinematic factors calculated from QED.  For parallel
and antiparallel spins $b$ is small allowing $g_1 (x,Q^2)$ to be measured
while if the target polarisation is perpendicular and parallel to the
lepton scattering plane $a$ is small allowing $g_2$ to be measured
(for the exact definitions of $a$ and
$b$ and the DIS variables $x$, $y$ and $Q^2$ see~\cite{KUHU}).
An alternative notation uses the virtual photon asymmetries which are given
by  $A_1 = (\sigma_{\frac{1}{2}}-\sigma_{\frac{3}{2}})/
(\sigma_{\frac{1}{2}}+\sigma_{\frac{3}{2}})$ and
$A_2 = \sigma_{TL}/\sigma_{T}$
where $\sigma_{\frac {1}{2}(\frac {3}{2})}$ is the photoabsorption
cross section in a state $J_z=\frac{1}{2}(\frac{3}{2})$
i.e. with the photon-nucleon spins antiparallel(parallel).
The cross-section $\sigma_{T}$ is the photoabsorption
cross section for transverse photons and $\sigma_{TL}$  is that due to the
transverse-longitudinal interference.  The asymmetry $A_1$ and $g_1$
are related by $A_1 = \sum_q  e_q^2 \left(q_R(x)-q_L(x)\right)/
\sum_q  e_q^2\left(q_R(x)+q_L(x)\right) \approx g_1/F_1$
where $F_1$ is the unpolarised structure function.
For parallel and antiparallel lepton and nucleon spins the measured
asymmetry, $A_m$, in any bin is related to $A_1$ and hence $g_1$ by:
\begin{equation}
A_m = \frac{N^{\uparrow\downarrow}-N^{\uparrow\uparrow}}
{N^{\uparrow\downarrow}+N^{\uparrow\uparrow}}  \approx P_B P_T f D A_1
\end{equation}
where $P_B$ and $P_T$ are the beam and target polarizations, the target
dilution factor, $f$, is the fraction of the events from the
free polarised nucleons in the target and the depolarization factor of
the virtual photon $D$ comes from QED and increases with $y$.
The product of the factors $P_B P_T f D$ is usually small so that
the asymmetries to be measured are also small.

In the early 1980s the SLAC experiments E80 and E130~\cite{SLAC}
reported the first measurements of polarised DIS for $x > 0.1$.
In 1988, the EMC reported measurements~\cite{EMC} over a range
down to $x = 0.015$. For $x > 0.1$
all the data agreed well with the expectations of the quark
parton model.  However, as $x$ decreased the EMC data fell progressively
below these expectations.

The significance of this disagreement with the quark parton model can be
understood by examining the first moment of the proton data which the EMC
measured to be
\begin{equation}
\Gamma_1^p = \int^1_0 g_1^p(x) dx = 0.126 \pm 0.01(stat) \pm 0.015(sys). 
\end{equation}
In the quark-parton model and to leading order in QCD this moment, 
for the proton and neutron, is given by
\begin{equation}
\Gamma_1^{p(n)} = +(-) \frac{1}{12} a_3 + \frac{1}{36} a_8
+ \frac{1}{9} a_0. 
\label{naiveg1}
\end{equation}
Here the $a_j$ are the diagonal combinations in the $SU(3)$ nonet 
of the axial matrix elements. With 
\be
\Delta q\,2M\,S^\mu = \langle P,S \vert \overline \psi \gamma^\mu 
\gamma_5 \psi \vert P,S\rangle   
\label{matrix}
\ee
where $S^\mu$ is the spin vector of the proton and the $a_j$ are 
related to the $\Delta q$ by 
$a_3 = g_A =\Delta u - \Delta d$,
$a_8 = (\Delta u + \Delta d -2 \Delta s)$ and
$a_0 = \Delta \Sigma = (\Delta u + \Delta d + \Delta s)$.
The matrix elements $a_3$ and $a_8$, measured from neutron and hyperon
decay, had values of 1.254 and 0.688, respectively, at the time of 
the EMC measurements.
Using these values and the measured value of $\Gamma_1^p$  implied
that $\Delta \Sigma = 0.12 \pm 0.09 \pm 0.14$ i.e.
only $\sim 12 \%$ of the spin of the proton is carried by its quarks.

This surprising result created great theoretical interest.  Where was the
spin of the nucleon ?  Could it be in the gluons ($\Delta G$)
as suggested in~\cite{EandT,AandR} or could it
be in orbital angular momentum ($L$)~\cite{[5.]}. By angular momentum
conservation the total spin of the nucleon of 1/2 must be equal to 1/2
$\Delta \Sigma +\Delta G + L$.  It was also
suggested that the problem did not exist and part of the measured moment
was missed in a perverse Regge behaviour which caused $g_1$ to diverge
in the unmeasured region at very small $x$~\cite{CandR}.

All this interest created a new experimental programme to investigate the
phenomenon further.  At SLAC experiments E142 and E143 were done at
incident polarized electron energies up to 29 GeV using polarized $^3$He,
hydrogenated and deuterated ammonia targets.  These were followed by
E154 and E155 at incident electron energies of 49 GeV using polarized
targets made from $^3$He, $^6$Li D and hydrogenated ammonia.  At DESY
the HERMES experiment uses polarized gas jet targets of $^3$He and
atomic hydrogen.  At CERN the Spin Muon Collaboration (SMC) has used
hydrogenated and deuterated ammonia and butanol polarized targets with
polarized muon beams of energies up to 200 GeV.  Such high energies allow 
coverage of a wide $x$ range. From all these experiments the 
structure function $g_1$ has been measured with much improved precision 
and the first measurements of the structure function $g_2$ have been 
reported.  The measurements of these structure functions for the 
proton and neutron provide the data which will be summarized in this talk.

The data from E142 and E143~\cite{E142E143} and the early data from 
SMC~\cite{SMC} confirmed the original EMC result~\cite{EMC} and provided
a precise and interesting check on the Bjorken sum rule~\cite{BJ}.
This is obtained from equation (5) by taking
the difference between the first moments for the neutron and the proton
so that in leading order in QCD; 
\begin{equation}
\Gamma_1^p - \Gamma_1^n = \frac {1}{6}(\Delta u - \Delta d) = \frac {g_A}{6}.
\label{bsr}
\end{equation}
This is a deeply fundamental sum rule which if broken implies serious
consequences for QCD~\cite{[10.]}.  The E142 and E143 data at
low $Q^2 \sim 2$ GeV$^2$ ~\cite{E142E143} are shown in Fig.~1.  
Before QCD radiative
corrections are applied the sum rule appears to be broken.  However,
as successive orders of QCD radiative corrections~\cite{CNS}~\cite{4order} 
(see section 3) are applied the data come into agreement with the sum rule.  
Hence, the Bjorken sum rule appears to be playing the same part in 
testing QCD as $g-2$ of the muon did for tests of QED.  We look 
forward to future more precise tests of QCD from this sum rule.
\begin{figure}[hb]
\begin{center}
\epsfig{file=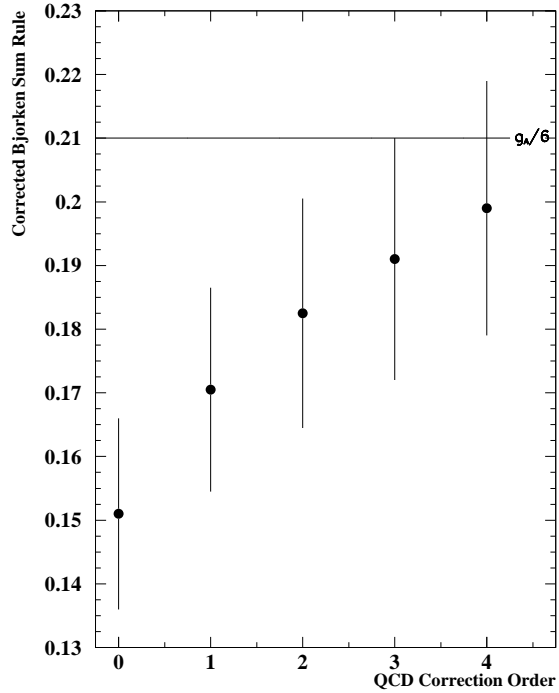,width=8cm}
\end{center}
\caption{
\label{fig1}\protect
The value of ($\Gamma_1^p-\Gamma_1^n)/C_{NS}$ where $C_{NS}$ 
is the QCD radiative correction defined in Eq.~\ref{ab} as 
a function of the order in $\alpha_s$ of the calculation of $C_{NS}$.  
The values of $\Gamma_1^p$ and $\Gamma_1^n$ are taken 
from~$^{10}$.
}
\end{figure}

\section{Data Reported at DIS 98}

\subsection{Measurements of $g_1$}

The SMC has presented data over the widest range of $x$ on the polarised
structure functions~\cite{[12.]}.  The collaboration has greatly improved
the precision of the data at low $x$ by demanding an observed hadron in 
each event.  This rejects radiative  
and other events with low depolarisation factors.  The remaining events
are then undiluted by data of poor significance for the asymmetry
determination allowing the asymmetry to be determined more precisely.
Furthermore, a much lower $Q^2$ trigger has been implemented which allows
asymmetries to be measured in the range $10^{-4} < x < 10^{-3}$.  The data
from this trigger serve to investigate the Regge region to search for a
possible divergence at low $x$ due to perverse behaviour such as that
proposed in~\cite{CandR}.  Fig.~2 shows the SMC data~\cite{Kiryluk}  with
the behaviour of $g_1 = 0.17/x\,\ln^2 x$ (solid curve) proposed
by~\cite{CandR}.  Such behaviour is excluded by the data.  
By implication even more extreme behaviours such as the power 
laws proposed in~\cite{Ryskinetal} will also be excluded but 
direct comparison is difficult due to the uncertainties in the 
hardness scale.  However, the 
less extreme behaviours $g_1 =-0.14\,\ln x$ (dashed curve) and 
$g_1 = -0.085(2+\ln x)$ (dotted curve) which were also proposed 
in~\cite{CandR} cannot be excluded.  All the curves 
were calculated assuming a value of $R=\sigma_L/\sigma_T=0$. Hence they 
represent lower limits since the curves scale as $1+R$.  The first of these 
behaviours would make some difference to the determination of 
$\Delta\Sigma$ so its exclusion removes a significant uncertainty.  
\begin{figure}[t]
\begin{center}
\epsfig{file=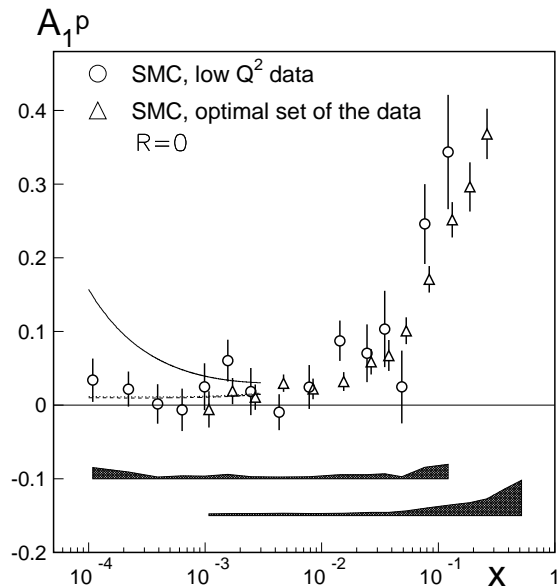,width=8cm}
\end{center}
\caption{
\label{fig2}\protect
The values of $A_1^p$ as a function of $x$ measured by 
SMC~$^{16}$.  
The smooth curves show the expected behaviour of $A_1^p$ 
as $x  \rightarrow 0$ proposed 
in~$^{7}$.  
The solid curve shows the behaviour for $g_1^p \sim 1/x\,\ln^2 x$, the 
dashed curve for $g_1 \sim \ln x$ and the dotted curve $g_1 \sim (2+\ln x)$.
}
\end{figure}

Data with impressive statistical precision were also shown by 
Hermes~\cite{[15.]} and E155~\cite{[16.]}.  The collaborations have not
yet evaluated the first moments over the full $x$ range but this will
be done soon so that we shall have more precise tests of QCD via the
Bjorken sum rule and improved measurements of $\Delta\Sigma$. 

\subsection {Measurements of $\Delta u_V$, $\Delta d_V$ and  $\Delta \bar q$
using inclusive hadrons}

These quantities were derived by both HERMES and the SMC from the
observed distributions of events in polarised DIS in which final
state hadrons were observed~\cite{[15.],[18.]}.  This method
assumes factorisation
in the quark parton model and measured values of the fragmentation
functions.  The integrals over the observed quark distributions are
consistent with those obtained from the structure function measurements.  
We look forward in the future to more precise measurements of these
quantities. In section 4, we will return to the use of semi-inclusive
measurements as a way to measure the gluon polarization.  Such 
measurements will probably be needed to determine how the nucleon's spin 
is shared between $\Delta G$ and $L$.  

\subsection{$g_2$ measurements}

E155 presented measurements~\cite{SRock} of $g_2$ which show consistency
with the Burkhardt-Cottingham sum rule~\cite{BC} and with earlier
SMC data~\cite{SMCg2}.  These measurements show that $g_2$ is small.
Hence it causes
little systematic error in the determination of $g_1$.  The data reported
came from a short run at SLAC.  A longer run which will allow much
improved precision is planned for next year. We will return to the
relevance of $g_2$ in section 5.2.

\section{Theoretical aspects of polarized DIS}

Next, we turn to some theoretical aspects that are important to 
understand in more detail the results in polarized DIS. Within the 
framework of QCD, one finds that a generic cross section or structure 
function $F(x,Q^2)$ can be written as a convolution
\be
F(x,Q^2) = \sum_i \int dy\,dz\,\delta(x-yz)\,C_i(y,Q^2/\mu^2,\alpha_s)
\,q_i(y,\mu^2),
\ee
where $q_i(y,\mu^2)$ are quark distribution functions at a given scale
(the soft part) and $C_i(y,Q^2/\mu^2,\alpha_s)$ are coefficient functions 
(the hard part) calculable in perturbative QCD. 
By taking moments of the structure functions, e.g. the lowest moment
$\Gamma(Q^2)$ = $\int^1_0 dx\,F(x,Q^2)$ one arrives at
\be
\Gamma (Q^2) = \sum_i C_i(Q^2/\mu^2,\alpha_s)\,q_i(\mu^2),
\ee
which relates the moments of structure functions via calculable 
coefficient functions $C_i(Q^2/\mu^2,\alpha_s)$ to moments of 
quark distributions, which can be expressed as scale dependent
matrix elements $q_i(\mu^2)$ of quark 
and gluon matrix elements between proton states. In Eq.~\ref{naiveg1}
we have already seen the above relation at leading order, 
$\alpha_s^0$, relating the first moment of $g_1(x,Q^2)$ to  
axial matrix elements $\Delta q$ or the combinations $a_i$. 
The `QCD-improved' result is given by
\bea
\Gamma_1^{p (n)}(Q^2) & = & +(-)\frac{1}{12}\,C_{NS}
\,\underbrace{\left( \Delta u - \Delta d\right)}_{a_3 (\rlap{$\bm\times$}\mu^2)}
+\frac{1}{36}\,C_{NS}\,\underbrace{\left( \Delta u + \Delta d 
- 2\Delta s\right)}_{a_8 (\rlap{$\bm\times$}\mu^2)}
\nonumber\\&&
+\frac{1}{9}\,C_{S}\,\underbrace{\left( \Delta u + \Delta d 
+ \Delta s\right)}_{\Delta\Sigma (\mu^2) = a_0 (\mu^2)}
\label{ms}
\eea
($\overline{MS}$ scheme), where $\rlap{$\bm\times$}\mu^2$ indicates
that the matrix element is scale-independent.
At order $\alpha_s^1$ (NLO) the coefficient functions $C_S$ =
$1-\alpha_s/3\pi + \ldots$ and $C_{NS}$ = $1-\alpha_s/\pi + \ldots$ 
%
%
and the next terms become scheme-dependent. 
The nonsinglet matrix elements are scale-independent, but
as a consequence of the nonconservation of the singlet axial current 
through the Adler-Bell-Jackiw anomaly, the matrix element $\Delta \Sigma
= a_0$, and hence also the $\Delta q$'s separately, are scale-dependent.  
Including a gluonic contribution in a redefinition of the matrix 
elements (distribution functions and their moments) one can define 
scale-independent quantities
\be
\Delta \tilde q(\rlap{$\bm\times$}\mu^2) = 
\Delta q(\mu^2) + \frac{\alpha_s}{2\pi}\,\Delta G(\mu^2),
\label{relation}
\ee
in which case one has
\bea
\Gamma_1^{p (n)}(Q^2) & = & +(-)\frac{1}{12}\,C_{NS}
\,a_3 (\rlap{$\bm\times$}\mu^2)
+\frac{1}{36}\,C_{NS}\,a_8 (\rlap{$\bm\times$}\mu^2)
\nonumber\\&&
+\frac{1}{9}\,C_{S}\,\Delta \tilde \Sigma (\rlap{$\bm\times$}\mu^2)
+ n_f\,C_G\,\frac{\alpha_s}{2\pi}\,\Delta G (\mu^2)
\label{ab}
\eea
(Adler-Bardeen scheme). The result for $\Gamma_1^p - \Gamma_1^n$ 
(Bjorken sum rule) only involves scale-independent matrix elements. It only
receives QCD corrections via the calculable coefficients $C_{NS}$, which
multiplies the result given in Eq.~\ref{bsr}. This coefficient is
now known~\cite{CNS} up to $\alpha_s^3$, with estimates~\cite{4order}  
for the next order 
and the calculations agree with experiment (see section 1, Fig.~1). 
The values of $\Gamma_1^p$ and $\Gamma_1^n$ each contain the scale-dependent 
matrix element $\Delta \Sigma(\mu^2)$ or $\alpha_s\,\Delta G(\mu^2)$
and hence they are useful as
measurements to determine this matrix element, but they should no longer
be referred to as sum rules~\cite{EJ}.

The most modern values of  $a_3$ and $a_8$ are 1.260 ~\cite{pdg} and  
0.575 ~\cite{FandD}, respectively.  Taken with 
$\Delta \Sigma$ (10 GeV$^2$) $\approx 0.25$, extracted from 
the measurements of 
$\Gamma_1^p$ and/or $\Gamma_1^n$, this gives us 
three pieces of data to learn about the quantities
appearing in Eqs.~\ref{ms} or \ref{ab}.   Three scenarios are sketched in 
which each differs in the value of $\Delta \tilde s$, the scale-independent
strange quark spin contribution. Choosing $\Delta \tilde s$ $\equiv$
0.00, -0.05 and -0.10 respectively, one sees in
Table~\ref{distr} what this implies for the scale-dependent polarized gluon
distribution, assuming $\alpha_s$ (10 GeV$^2$)
$\approx$ 0.25.
\begin{table}
\caption{\label{distr} \protect Values for the scale-independent quark spin
contributions $\Delta \tilde q$ (see Eq.~\ref{relation}) for different
scenarios for the strange quark matrix element $\Delta \tilde s$.
Quantities in parentheses are fixed by experiments.}
\begin{center}
\begin{tabular}{|l|c|c|c|}
\hline
Quantity & Scenario & Scenario & Scenario \\
& I & II & III \\
\hline
$\Delta \tilde u (\rlap{$\bm\times$}\mu^2)$ &
+0.92 & +0.87 & +0.82 \\
$\Delta \tilde d (\rlap{$\bm\times$}\mu^2)$ &
-0.34 & -0.39 & -0.44 \\
$\Delta \tilde s (\rlap{$\bm\times$}\mu^2)$ &
-0.00 & -0.05 & -0.10 \\
\hline
$a_3(\rlap{$\bm\times$}\mu^2)$ & (1.26) & (1.26) & (1.26) \\
$a_8(\rlap{$\bm\times$}\mu^2)$ & (0.58) & (0.58) & (0.58) \\
\hline
$\Delta \tilde \Sigma(\rlap{$\bm\times$}\mu^2)$ & 
0.58 & 0.43 & 0.28 \\
\hline
$\Delta \Sigma$ (10 GeV$^2$) & 
(0.25) & (0.25) & (0.25) \\
$\Delta G$ (10 GeV$^2$) & 2.8 & 1.5 & 0.3 \\
\hline
\end{tabular}
\end{center}
\end{table}
The strong variation of $\Delta G$ (10 GeV$^2$) shows the insensitivity of
inclusive DIS for this quantity. The results of a more elaborate analysis
for the polarized quark and gluon distributions in both the $\overline{MS}$ 
and AB schemes were presented in the talk by Sichtermann ~\cite{[14.]}.  
Using the different schemes, the 
results (omitting systematic errors) for $\Delta G$ were found to be 
\bea
&&
\Delta G (5\ \rm{GeV}^2) = 0.25^{+0.29}_{-0.22} 
\qquad (\overline{MS}\ \mbox{scheme})
\\ &&
\Delta G (5\ \rm{GeV}^2) = 0.99^{+1.17}_{-0.31}. 
\qquad (\mbox{AB scheme})
\eea
This illustrates the insensitivity of inclusive DIS to the polarized gluon
distribution. If in a more precise analysis of this type the scheme dependence
persists it will be an indication that NNLO effects need to be included in 
this type of analysis. We also note that at this stage of refinement,
i.e. incorporating DGLAP evolution, 
approximations such as $A_1(x,Q^2) \approx A_1(x)$ should not
be used in the data analysis (see also
Ref.~\cite{kotikov}). 

Returning to the spin of the nucleon and the sum rule
$1/2$ = $1/2\,\Delta \Sigma + \Delta G + L_q + L_G$,
we see that the inclusive analysis shows us the importance of an
{\em independent} measurement of $\Delta G$. Furthermore, we know that
$\Delta G(Q^2) \propto 1/\alpha_s(Q^2)$ and so it grows with 
increasing $Q^2$. This points to the importance of measuring the orbital 
contribution to the angular momentum~\cite{Ji} as well as $\Delta G$. 
One possibility for this which is being investigated is to use 
virtual Compton scattering~\cite{DVCS} 
($\gamma^\ast + p \rightarrow \gamma + p^\prime$). 
However, the cross section for such an exclusive process is rather small.

\section{Plans to measure $\Delta G$}

In inclusive DIS, the measurements of the polarized gluon distribution
is indirect. It enters via the evolution equations, as discussed in the
previous section. Only a wide $Q^2$
range of the data could improve our knowledge of $\Delta G$, possibly
requiring theoretical efforts to go to sufficiently high order if the
problem of the scheme dependence of the NLO QCD persists~\cite{[14.]}.
The option of a future polarized $\vec e\vec p$ HERA collider, which 
requires the polarization of protons in the HERA ring, would provide 
an excellent opportunity for such a measurement~\cite{Radel}.

A more direct way to measure the gluon distribution is the photon-gluon
fusion process, $\gamma^\ast + G \rightarrow q\bar q$. This is planned
in the HERMES experiment~\cite{[15.]} via the fusion into a $c\bar c$ pair,
which is reconstructed from the decay of charmed mesons in the final state.
Rather than determining the sum rule the experiment is sensitive to $\Delta
G(x)/G(x)$ at $x \approx 0.3$. This approach is also proposed
by the COMPASS~\cite{Mallot} experiment, which is sensitive 
to $\Delta G(x)/G(x)$ at 
$x \approx 0.1$. Another option to obtain the polarized gluon distribution 
is being studied by COMPASS.  This consists of detecting the production of
a light $q\bar q$ pair from photon-gluon fusion via tagging of an 
oppositely charged pair of hadrons with large transverse momentum. 
At a future polarized HERA collider the photon-gluon 
fusion into a pair of jets is the obvious final state tag. This would
give access to $\Delta G(x)$ for $x \ge 0.0015$. A complication in this 
case is the Compton process $\gamma^\ast + q \rightarrow q + G$, which also
leads to a two jet final state. Important to note is the existence of
the calculations of the NLO QCD corrections
for charm production, reported at this meeting~\cite{[19.]}.

Hadron induced processes at RHIC~\cite{Hayashi} and HERA-$\vec N$, the option
using polarized protons in HERA in combination with a gas jet 
target~\cite{Radel} were also discussed at this meeting. On the theoretical
side, experimentalists were warned of other possible small-$x$ complications 
besides the effects of gluons~\cite{Badelek,Kochelev}.

\section{Unravelling the nucleon structure}
\subsection{Nonlocal quark and gluon operators}
The moments of structure functions are related via the operator product 
expansion to local quark and gluon operator combinations. The $x$-dependent
quark and gluon distributions themselves also can be expressed in terms
of quark and gluon operators, albeit that they are no longer local. One
has for the polarized quark helicity distributions,
\be
\Delta q(x) = \left.  \int \frac{d\xi^-}{4\pi}\ e^{i\,xP^+\xi^-} 
\,\langle P,S\vert \overline\psi(0)\gamma^+\gamma_5\,\psi(\xi)\vert
P,S\rangle\right|_{\xi^+ = \xi_T = 0},
\ee
which is a lightcone correlation function\footnote{
$a^\pm = (a^0 \pm a^3)/\sqrt{2}$ are lightcone components},
and for the polarized gluon distribution one has
\be
x\Delta G(x) = \left.  \int \frac{d\xi^-}{4\pi}\ e^{i\,xP^+\xi^-} 
\,\langle P,S\vert F^{+\alpha}(0)\,\tilde F^{+}_{\ \alpha}(\xi)\vert
P,S\rangle\right|_{\xi^+ = \xi_T = 0}.
\ee
For instance one immediately recovers the result in Eq.~\ref{matrix}
after integrating the above result for $\Delta q(x)$ over $x$. 
The knowledge of the quark and gluon distributions as matrix 
elements enables
\begin{itemize}
\item
Interpretation of quantities as densities, which requires rewriting
in terms of good (``partonic") components of the quark field,
$\psi_+ \equiv 1/2\,\gamma^-\gamma^+\,\psi$. One has
\begin{eqnarray}
&&
\overline{\psi}\gamma^+\psi \propto \psi_{+}^\dagger \psi_{+},
\label{opa}
\\
&&
\overline{\psi}\gamma^+\gamma_5\psi \propto \psi_{+R}^\dagger \psi_{+R}
-\psi_{+L}^\dagger \psi_{+L},
\label{opb}
\end{eqnarray}
where 
$\psi_{+R/L} \equiv 1/2\,(1\pm \gamma_5)\,\psi_+$ are chiral projections.
\item
Study of scale dependence (evolution). For more details on this we
refer to the working group on structure functions.
\item
Calculation of the distributions. Lattice calculations can be used to
obtain local matrix elements. Explicit models for hadrons, such as
quark potential models, bag models, etc. in principle
can be used, given the structure of the matrix element. A difficulty 
here is the usual lack of a fully covariant description.
\end{itemize}

\subsection{Higher twist contributions: $g_2$}

The remarks in the previous paragraph are perhaps best illustrated
on the example of the distribution $g_T(x)$ = $g_1(x) + g_2(x)$, 
which is the $1/Q$-suppressed combination in the asymmetry 
in Eq.~\ref{asym}. It is connected to ``less-trivial" operators than 
the one in the previous section, in this case
\be
S_T^\alpha\,g_T(x) =
\left.  \int \frac{d\xi^-}{4\pi}\ e^{i\,xP^+\xi^-} 
\,\langle P,S\vert \overline\psi(0)\gamma^\alpha\gamma_5\,\psi(\xi)\vert
P,S\rangle\right|_{\xi^+ = \xi_T = 0},
\ee
After some massaging, using Lorentz covariance and the QCD
equations of motion $g_2$ can be separated into two parts,
\be
g_2(x) = g_2^{WW}(x) + \tilde g_2(x),
\ee
where the first term is the Wandzura-Wilczek contribution~\cite{WW}, 
which can
be calculated from $g_1$ and a so called interaction-dependent part,
which involves quark-gluon-quark correlation functions. 
For $g_2$ several sum rules exist, the most well-known being the
Burkhardt-Cottingham sum rule~\cite{BC}, $\int dx\ g_2(x) = 0$, which
is also satisfied by the separate terms in the above relation for $g_2$. 
Other sum rules that recently
have been studied are the Efremov-Leader-Teryaev sum rule~\cite{ELT}.
The new E155 data, described in section 2.3, show consistency with
the Burkhardt-Cottingham sum rule and indicate that $\tilde g_2 (x)$
is small.
In Ref.~\cite{SRock} a comparison of the particular moment $d_2$ with
theoretical estimates from models and lattice calculations has been
shown.

\subsection{Transverse spin distribution: $h_1$ or $\Delta_T q$}

At leading order the soft part governing the quark content of a spin 1/2
hadron can be described with three quark distributions. The most well-known
are the unpolarized quark distribution $q(x)$ or $f_1(x)$
and the quark chirality (or helicity) distribution $\Delta q(x)$ or $g_1(x)$, 
involving the operator structures in Eqs.~\ref{opa} and \ref{opb} respectively.
The third one is the quark transverse spin distribution $\Delta_Tq(x)$ or
$h_1(x)$. 
While $g_1(x)$ involves chiral projections $\psi_{+R/L}$ of the quark field, 
the function $h_1(x)$ contains the following quark operators,
\be
\overline{\psi}\gamma^1\gamma^+\psi \propto
\psi_{\uparrow}^\dagger \psi_{+\uparrow},
-\psi_{+\downarrow}^\dagger \psi_{+\downarrow},
\ee
where
$\psi_{+\uparrow/\downarrow} \equiv 1/2\,(1\pm\gamma^1\gamma_5)\,\psi_+$
are transverse spin projections. It requires a transversely polarized
nucleon to measure this function. Rewritten in terms of chiral quark 
fields, the operator structure is $\psi_{+R}^\dagger \psi_{+L}$,
hence the soft part requires a chirality flip and the function is 
referred to as {\em chirally odd}. In hard processes, this can only
be achieved by a nonzero quark mass, hence the function is suppressed in
inclusive deep inelastic scattering. In semi-inclusive processes, it is
possible to combine two soft chirally odd parts, the first describing the 
quark content of the target and the second describing the quark fragmentation 
into hadrons. We mention two possibilities,
\begin{itemize}
\item
Drell-Yan scattering~\cite{Kumano}, where the asymmetry~\cite{JJ}
\be
A_{TT} \propto \sum_a h_1^a(x_{\sst A})\,h_1^{\bar a}(x_{\sst B}),
\ee
is sensitive to annihilating transversely polarized quarks in the
scattering of two transversely polarized nucleons. 
\item
Semi-inclusive DIS, where one can consider spin transfer
from a transversely polarized nucleon via a transversely polarized
quark to a transversely polarized hadron in the final state, e.g. the
process~\cite{Artru} $ep^\uparrow \rightarrow e\Lambda^\uparrow X$. Another 
possibility~\cite{Jakob} is to use azimuthal asymmetries in semi-inclusive
DIS, e.g.
\be
\langle \sin(\phi_h^\ell+\phi_S^\ell)\rangle_{OT}
\propto \sum_a h_1^a(x)\,H_1^{\perp (1) a}(z),
\ee
where the function $H_1^{\perp (1) a}(z)$ is a fragmentation function
into an unpolarized or spinless final state, which correlates the
quark transverse spin to the transverse momentum relative to 
the produced hadron.  In this way it leads to an azimuthal asymmetry
involving the azimuthal angles $\phi_h^\ell$ and $\phi_S^\ell$ of a 
produced hadron or a target spin relative to the lepton scattering 
plane~\cite{Collins,lh}.
\end{itemize}
The evolution of these functions is known and is different from that of 
the helicity distributions~\cite{Kumano}.

\section{Conclusion}

Several other items have been discussed in the spin working group. We
have just mention the detailed investigation of the effects of transverse
quark momenta in various hard processes such as 
 Drell-Yan scattering, semi-inclusive
DIS and $e^+e^-$ annihilation~\cite{Jakob,lh,hard}, the quark distributions in
higher spin targets and their sensitivity to specific correlations in
the target, e.g. the presence of pionic components~\cite{Schaefer}.
We also mention the possibilities to use polarization to zoom in on
new phenomena~\cite{Radel}, e.g. to decide on contact interactions 
proposed~\cite{Virey} to explain the possible excess of 
high $Q^2$ events at HERA, or
to improve our knowledge of the polarized quark distributions via
charged current interactions. Both of these are examples of new
possibilities offered by polarized protons in HERA. This also offers
possibilities to study spin effects in diffractive processes~\cite{Kop}
or to study the polarized quark distributions in the photon~\cite{Gehr}.

Returning to the present,  at this meeting we have seen 
measurements of impressive precision of the polarised
structure functions $g_1$ and $g_2$ of the proton and neutron.  We look
forward to further data in the future.  Evidence has been presented 
in the workshop which shows that the observed deficiency of the spin 
of the nucleon carried by the quarks is not due to perverse Regge 
behaviour which affects the extrapolation to small $x$.  Hence the 
problem of the distribution of the nucleon's spin still persists. 
It is still not possible to say whether the deficiency of the 
nucleon's spin carried by the quarks is taken up by
$\Delta G$ or $L$ and theoretical refinements will be required here
to understand the scheme dependence of the QCD fits to the data.  
We look forward to future experiments (e.g. HERMES, Compass, RHIC, 
E156, HERA) which will make direct determinations of $\Delta G$.

\section*{Acknowledgments}
We thank the organisers for creating such an interesting, stimulating
and enjoyable workshop.  We also thank all the participants in Working Group
4 for their assistance in preparing this talk.  We are grateful to R G Roberts 
for valuable discussion.

\end{document}